\documentclass[preprint]{aastex}

\newcommand{\arcm}{\ifmmode {' }\else $' $\fi}
\newcommand{\arcs}{\ifmmode {'' }\else $'' $\fi}
\slugcomment{}

\shortauthors{Rhode \& Zepf}
\shorttitle{Globular Cluster System of NGC 7814}

\begin{document}

\title{The Globular Cluster System of the Spiral Galaxy NGC~7814}

\author{Katherine L. Rhode\altaffilmark{1}}
\affil{Department of Astronomy, Yale University, New Haven, CT 06520}
\email{rhode@astro.yale.edu}

\author{Stephen E. Zepf}
\affil{Department of Physics \& Astronomy, Michigan State University, East Lansing, MI 48824}
\email{zepf@pa.msu.edu}

\altaffiltext{1}{NASA Graduate Student Researchers Program Fellow}

\begin{abstract}
We present the results of a wide-field photometric study of the
globular cluster (GC) system of the edge-on Sab spiral NGC~7814.  This
is the first spiral to be fully analyzed from our survey of the GC
systems of a large sample of galaxies beyond the Local Group.
NGC~7814 is of particular interest because a previous study estimated
that it has 500$-$1000 GCs, giving it the largest specific frequency
($S_N$) known for a spiral.  Understanding this galaxy's GC system is
important in terms of our understanding of the GC populations of
spirals in general and has implications for the formation of massive
galaxies.  We observed the galaxy in BVR filters with the WIYN 3.5-m
telescope, and used image classification and three-color photometry to
select GC candidates.  We also analyzed archival HST WFPC2 images of
NGC~7814, both to help quantify the contamination level of the WIYN GC
candidate list and to detect GCs in the inner part of the galaxy halo.
Combining HST data with high-quality ground-based images allows us to
trace the entire radial extent of this galaxy's GC system and
determine the total number of GCs directly through observation.  We
find that rather than being an especially high-$S_N$ spiral, NGC~7814
has $_<\atop{^\sim}$200 GCs and $S_N$ $\sim$ 1, making it comparable
to the two most well-studied spirals, the Milky Way and M31.  We
explore the implications of these results for models of the formation
of galaxies and their GC systems.  The initial results from our survey
suggest that the GC systems of typical ellipticals can be accounted
for by the merger of two or more spirals, but that for highly-luminous
ellipticals, additional physical processes may be needed.

\end{abstract}

\keywords{galaxies: individual (NGC~7814); galaxies: star clusters;
galaxies: spiral}

\section{Introduction}

Globular clusters (GCs) --- with their old ages, extended spatial
distributions and uniform chemical compositions --- are powerful tools
with which to study the formation of galaxies.  Since the late 1980s,
the study of GC systems of external galaxies has developed into a
full-fledged subfield of astronomy (see Ashman \& Zepf 1998, Harris
2001), with most of the emphasis placed on using GCs to piece together
the dynamical and chemical enrichment history of giant galaxies.  Much
of the observational effort has focused on the GC systems of
elliptical galaxies, including detailed investigations of their inner
regions with the Hubble Space Telescope (e.g., Kundu \& Whitmore 2001,
Larsen et al.\ 2001) and some new wide-field studies using mosaic CCD
detectors on ground-based telescopes \citep{rhode01,dirsch03}.  But
studies of the GC systems of spirals beyond the Local Group have
lagged considerably behind those of ellipticals.  A few HST studies of
spirals have been done (Kissler-Patig et al.\ 1999, Goudfrooij et al.\
2003), but their limited spatial extent prevents them from addressing
the global properties of the galaxies' GC systems.  Most previous
ground-based studies were carried out many years ago using either
small-format CCDs or photographic plates.

As a result, fundamental issues remain about the comparison between
the GC systems of spirals and ellipticals, due in large part to the
fact that our knowledge of spiral galaxy GC systems depends almost
solely on the Milky Way and M31.  For example, although there is
reasonable evidence that elliptical galaxies in general have more GCs
per unit mass than spirals, how large this difference is and how it
depends on properties like morphological type and environment of the
host galaxy are uncertain (see, e.g., Ashman \& Zepf 1998,
Kissler-Patig, Forbes, \& Minniti 1998).  The question of why
ellipticals have more GCs per unit mass or luminosity than spirals has
motivated much of the work in this field, but further progress depends
on establishing the detailed properties of the GC systems of {\it
both} ellipticals and spirals.

In an attempt to remedy some of the weaknesses in the census of spiral
GC systems, we are carrying out a wide-field survey of the GCs around
a good-sized sample of spirals beyond the Local Group.  Our overall
objectives are to expand the database of well-studied spirals and to
use the information we gather to gain insight into how galaxies form.
One specific aim is to test the predictions of formation scenarios for
ellipticals.  For example, prompted in part by evidence that
elliptical GC systems are more populous and have higher mean
metallicity than those of spirals, Ashman \& Zepf (1992; hereafter
AZ92) suggested that ellipticals form via mergers of two or more disk
galaxies and that GCs are formed in these mergers.  One of the
predictions of AZ92 is that if Es form in galaxy mergers, they should
have GC systems with bimodal color distributions; this was later
confirmed by observations (e.g., Zepf \& Ashman 1993, Gebhardt \&
Kissler-Patig 1999, Kundu \& Whitmore 2001, Larsen et al.\ 2001).
Another consequence of the AZ92 model is that the mass-normalized
number of blue, metal-poor GCs in spiral galaxies should be the same
as in ellipticals. This is because the blue, metal-poor GC population
comes directly from the progenitor disk galaxies and the red,
metal-rich GCs are formed in the merger.  More broadly, any difference
in the number of metal-poor GCs as a function of galaxy type or
environment would provide important clues to differences in the
physical conditions in galaxies at early stages of their formation.
These ideas can only be tested by establishing what these numbers are
for a sample of spirals of different types.

Accordingly, we have obtained imaging data in three broadband filters
for nine spiral galaxies, ranging in morphological type from Sab to
Sc.  All the targets are close to edge-on (with inclinations
$_<\atop{^\sim}$80 degrees) so that the GCs can be more easily
detected, and all have distances of 20 Mpc or less.  Most have data
available in the HST archive, which we use to help characterize
background galaxy contamination in the GC samples.  As part of the
same survey, we have also observed a sample of early-type galaxies out
to large galactocentric radii in order to better constrain their GC
system properties.  A brief description of the overall study
(ellipticals and spirals) is given in Rhode \& Zepf (2001; Paper~I).

We chose to begin our investigation of spirals with the Sab spiral
NGC~7814.  This galaxy was studied by Bothun, Harris, \& Hesser~(1992;
hereafter B92), who detected $\sim$120 GC candidates and estimated
that the total number around the galaxy is $\sim$500$-$1000.  B92
calculated a final specific frequency ($S_{N}$, the total number of
GCs per unit luminosity of the host galaxy in units of $M_{V}$ $=$ 15)
between 5.2$\pm$2.2 and 7.6$\pm$3.2, depending on the assumed distance
to NGC~7814.  These are very large numbers compared to the Milky Way
and M31, which have $S_N$ $\sim$ 0.5$-$0.7 and 0.9$-$1.3, respectively
\citep{az98,barmby03}, and make NGC~7814 an obvious candidate for
further study.

Here we present our findings for the GC system of NGC~7814.  The next
section describes our observations and initial data reduction steps.
Section~\ref{section:analysis} describes the analysis of both the
ground-based and HST data, and Section~\ref{section:results} gives the
results.  The paper ends with a discussion and summary of our results
and their implications for spiral galaxy GC systems and galaxy
formation models.

\section{Observations and Initial Reductions}
\label{section:obs and redux}

Images of NGC~7814 were obtained in 1998 December and 1999 October
with the 3.5 m WIYN telescope\footnote{The WIYN Observatory is a joint
facility of the University of Wisconsin, Indiana University, Yale
University, and the National Optical Astronomy Observatory.} at Kitt
Peak National Observatory.  The detector used was a 2048x2048 CCD
(S2KB), which has 0.196$\arcs$-pixels and a field-of-view 6.7$\arcm$
on a side when mounted on WIYN.  To observe as much of the galaxy halo
and GC system as possible, we positioned the center of NGC~7814 in one
corner of the CCD, which gives us radial coverage to $\sim$9$\arcm$,
or $\sim$35 kpc.  To help separate {\it bona fide} GCs from
contaminating objects (foreground stars and background galaxies),
images were obtained in three broadband filters (BVR).  Multiple
exposures were taken in each filter and the telescope was dithered
between exposures to facilitate cosmic ray removal.  Total integration
times were 7200~s (four exposures) in $B$, 5400~s (three exposures) in
$V$ and 5400~s (three exposures) in $R$.  One $B$-band, one $V$-band,
and the three $R$-band frames were taken under photometric conditions,
and the rest of the images were taken on mostly clear, but not
photometric, nights.
Standard star fields (Landolt 1992) were observed during one night in
1999 October for use in the photometric calibration.  The errors on
the zero-point constants ranged from 0.003 to 0.005 magnitudes,
indicating that the night was indeed photometric.  

Preliminary reductions (overscan and bias level subtraction,
flat-field division) of the WIYN images were accomplished using
standard IRAF\footnote{IRAF is distributed by the National Optical
Astronomy Observatories, which are operated by AURA, under cooperative
agreement with the National Science Foundation.} tasks.  Sky
subtraction was performed on the individual images taken in each
filter before they were scaled to a common flux level and combined.
The background level was restored to the combined images and they were
aligned and flipped to a north-up, east-left orientation.  The
resolution (point-spread function FWHM) of the combined images is
1.0$\arcs$ in B, 1.0$\arcs$ in V, and 1.3$\arcs$ in R.

In addition to the WIYN data, we also made use of data from the HST
archive\footnote{Based on observations made with the NASA/ESA {\it
Hubble Space Telescope}, obtained from the data archive at the Space
Telescope Science Institute.  STScI is operated by AURA, under NASA
contract NAS 5-26555.} for this study.  Two Wide-Field and Planetary
Camera 2 (WFPC2) data sets were available in the archive.  Images from
program GO.8597 (PI: Regan) were positioned with the center of the
galaxy on the PC chip, and images from program GO.6685 (PI: Huizinga)
were positioned away from the galaxy center, to include more of the
halo.  The former, which we will call the galaxy pointing, consisted
of two exposures in the F606W filter of length 160~s and 400~s.  The
latter data set, hereafter referred to as the halo pointing, consisted
of two 400~s frames in F450W, one 600~s frame in F555W, and one 600~s
frame in F814W.  ``On-the-fly'' calibration was applied to the data
sets when we requested them from the archive.  Multiple images of a
given pointing taken in the same filter 
were combined using the STSDAS task CRREJ.  The HST data were used to
help quantify the contamination level of the WIYN data, as well as to
find GCs in regions close to the galaxy disk; further analysis of the
HST images is described in Sections~\ref{section:contamination} and
\ref{section:hst sources}.

\section{Data Analysis}
\label{section:analysis}

The techniques we use to detect, select, and analyze GC candidates
around our target galaxies are explained in detail in Paper~I. Here we
describe the basic analysis steps for NGC~7814; the reader should
consult Paper~I for additional information.

\subsection{Source Detection}
To remove the diffuse galaxy light and detect discrete sources around
NGC~7814, the WIYN images were first smoothed with a circular median
filter with a diameter 7 times the mean FWHM of point sources in the
image.  The smoothed image was subtracted from the original and a
constant sky level was added to restore the background.  The IRAF task
DAOFIND was used to detect sources above a given signal-to-noise
threshold in the galaxy-subtracted image.  A total of 398 objects were
detected that appear in all three filters.

\subsection{Extended Source Cut}
\label{section:extended source cut}
\begin{figure}
\plotone{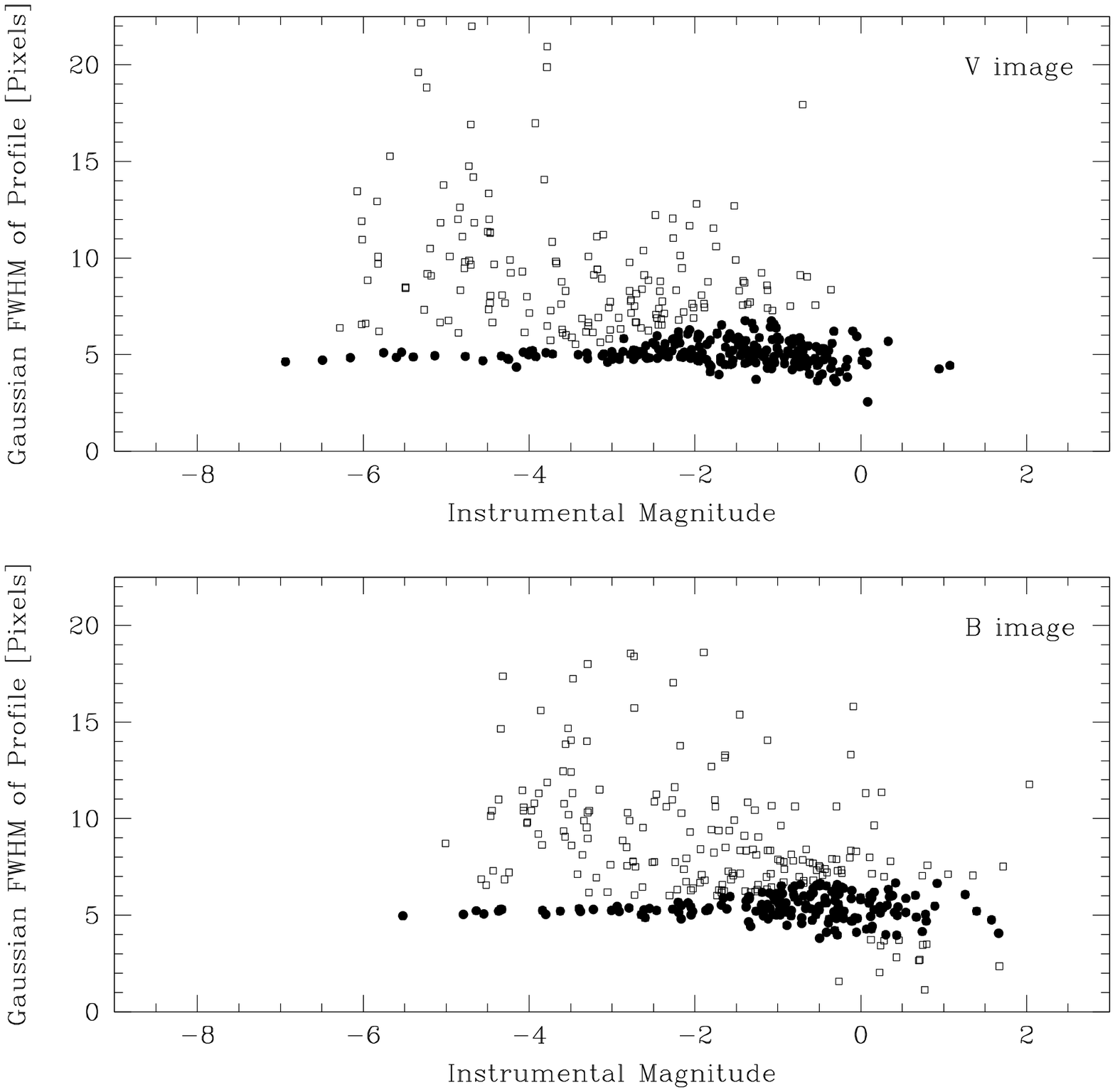}
\caption{Gaussian FWHM of the radial profile versus instrumental
magnitude for the 398 detected objects in the WIYN $V$ and $B$ images.
Filled circles are objects that passed the extended source cut and
open squares are objects deemed extended and therefore eliminated from
the list of possible GCs.  \protect\label{fig:fwhm mag}}
\end{figure}

An extended source cut was applied to the WIYN source list in order to
remove as many background galaxies as possible from the GC candidate
list.  The quality of the $B$ and $V$-band images was significantly
better than that of the $R$ image,
so the latter was not used for this step.  The FWHM values of the 398
detected sources were measured in the $B$ and $V$ images and plotted
versus their instrumental magnitudes.  Figure~\ref{fig:fwhm mag} shows
FWHM versus instrumental magnitude for the 398 objects in the $B$ and
V images, with filled circles indicating objects that met our
selection criteria for point sources and open circles marking objects
that were deemed extended and thus eliminated.  As with NGC~4472
(Paper~I), the HST images were used to fine-tune the extended source
cut at the faint end.  We eliminated 228 objects that were extended in
either the $B$ or $V$ frames, leaving 170 apparent point sources in
the WIYN images.  This is a relatively large proportion of extended
objects (60\%); only 20\% of the sources were extended in the NGC~4472
study.  Indeed, upon examination this excess of extended objects is
immediately apparent in the WIYN images: there are a large number of
faint background galaxies across the entire frame, forming what looks
like a cluster or group of galaxies behind NGC~7814.

\subsection{Photometry}
Sixteen standard stars from four Landolt fields observed during the
photometric night in 1999 October were used to derive photometric
calibration coefficients (zero points and color terms).  Photometry in
an aperture of radius equal to the mean FWHM of the image was carried
out in $B$, $V$, and $R$ on the 170 objects that passed the extended
source cut. Calibrated total magnitudes were then calculated using
aperture corrections and the photometric coefficients. The aperture
corrections, which were derived in the same manner as for Paper~I, are
listed in Table~\ref{table:apcorr}.  A galactic extinction correction,
taken from the reddening maps of \citet{schlegel98}, was applied to
the final magnitudes; the corrections in the direction of NGC~7814 are
$A_B$ $=$ 0.194, $A_V$ $=$ 0.149, and $A_R$ $=$ 0.120.

\subsection{Color Selection}
\label{section:color cut}
The final step in the GC candidate selection is to choose the subset
of point sources with magnitudes and colors expected for GCs.  We
assumed that the brightest GC would have $M_V$ $=$ $-$11, based on the
GCLF of the Milky Way and other galaxies \citep{az98}.  The distance
to NGC~7814 has recently been determined via the surface brightness
fluctuation (SBF) technique by \citet{tonry01}, who find $m-M$ $=$
30.60$\pm$0.14.  We adopt this distance for the rest of our
calculations, unless noted otherwise.  This yields an expected
magnitude of $V$ $=$ 19.6 for the brightest GC around NGC~7814.

To select GC candidates by color, we used the McMaster catalog
of Galactic GCs \citep{harris96} to derive a linear relationship
between $B-V$ and $V-R$ color for GCs.  We then chose a range of $B-V$
color for the selected GC candidates.  The chosen range is 0.56 to
0.99, which corresponds to [Fe/H] between $-$2.5 and 0.0 for Galactic
GCs.  Our selection algorithm is as follows: objects with $V$ fainter
than 19.6 and photometric errors in all three filters of less than
0.15~mag were retained. Then, taking into account their individual
photometric errors, objects with $B-V$ and $V-R$ that put them within
2.5-sigma of the derived relation for Galactic GCs were accepted as GC
candidates.  This yielded a list of 42 sources.

\begin{figure}
\plotone{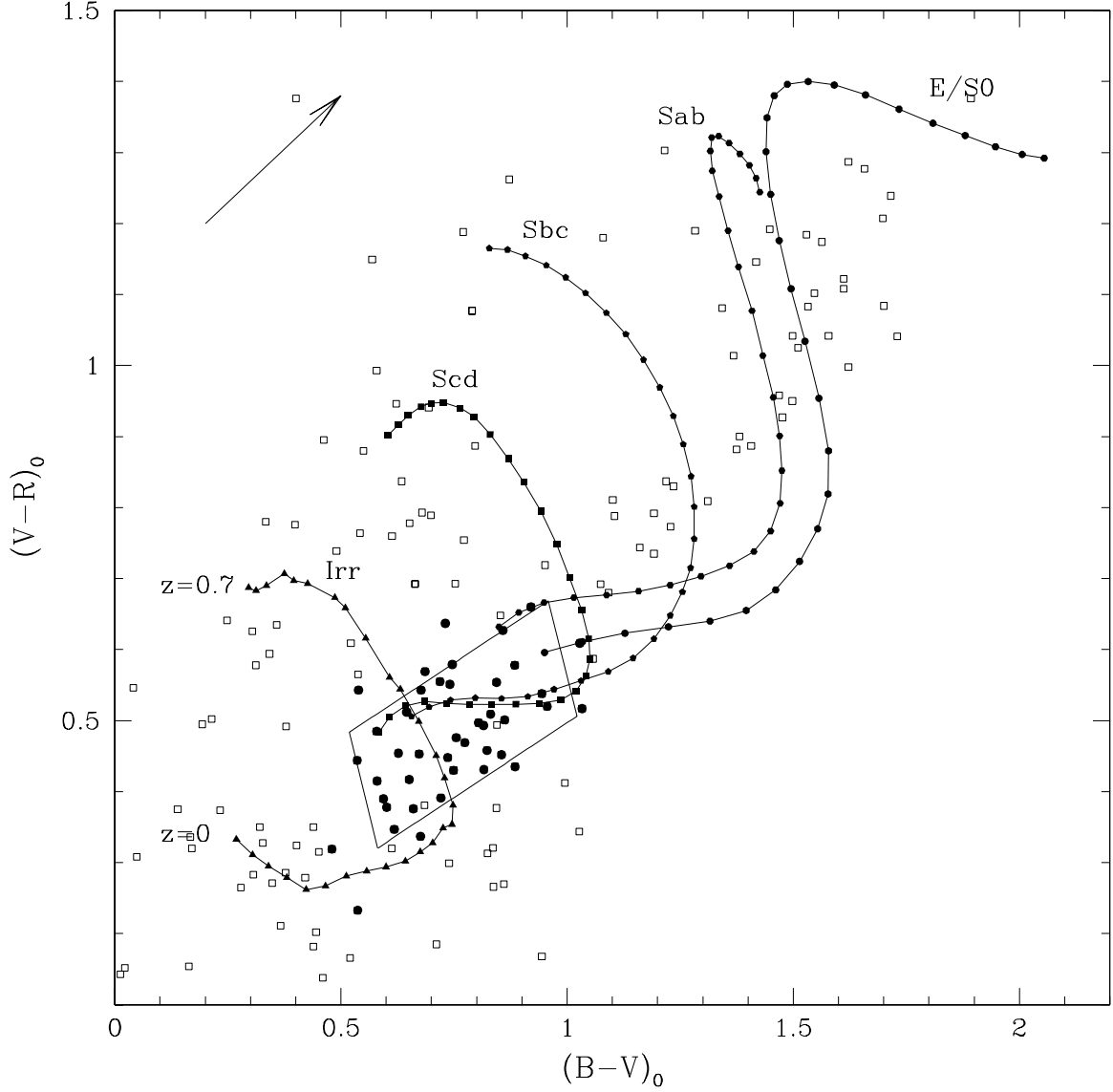}
\caption{Color selection of GC candidates around NGC~7814. The colors
of the 170 objects that appear as point sources in the WIYN images are
shown here.  Open squares are objects that failed to meet the
magnitude/color selection criteria and filled circles are the final
set of 42 GC candidates.  For reference, the locations in the
color-color plane of galaxies of various types are shown as tracks the
galaxies would follow with increasing redshift.  A reddening vector of
length $A_V$ $=$ 1~magnitude appears in the upper left corner.
\protect\label{fig:color cut}}
\end{figure}

Figure~\ref{fig:color cut} shows the results of the color selection.
All 170 point sources in the WIYN image are plotted in the BVR
color-color plane, with rejected objects as open squares and GC
candidates as filled circles.  For illustrative purposes, the
locations of galaxies of various types are shown as ``tracks'' the
galaxies would follow in the BVR plane with increasing redshift.
Paper~I describes how the tracks were produced.  GCs coincide in the
BVR plane with late-type, low- to moderate-redshift galaxies, some of
which will have been eliminated from the GC sample by the extended
source cut. Three-color photometry enables us to eliminate significant
numbers of contaminating objects from the GC candidate list, but some
contamination still remains.  Section~\ref{section:contamination}
describes how we quantify the contamination level.

\subsection{Completeness Testing}
\label{section:completeness}
A series of completeness tests was carried out to establish the point
source detection limit in the WIYN images. Fifty artificial point
sources with magnitudes within 0.1 magnitude of a given brightness
were added to each image, the same detection steps used on the
original image were performed, and the fraction of artificial sources
detected was recorded.  Fifty to sixty such tests were executed on
each image so that the completeness was calculated over a range of 5
to 6 magnitudes per filter.  The data are 50\% complete at $B$ $=$
25.3, $V$ $=$ 24.8, and $R$ $=$ 23.9.

\subsection{Quantifying and Correcting for Contamination}
\label{section:contamination}

\subsubsection{Galaxies}
\label{section:galaxy contamination}

HST resolves many faint background objects that can appear as point
sources in ground-based images, so to estimate the galaxy
contamination in the selected GC sample, we analyzed the archival HST
data described in Section~\ref{section:obs and redux}.  Sixteen of the
42 WIYN GC candidates were located in at least one of the two HST
pointings.  We followed the method of \cite{kundu99} to determine
which of these were true point sources.  Photometry was performed with
aperture radii of 0.5 pixels and 3 pixels and a sky annulus from 5 to
8 pixels.  Objects in the PC chip with
counts$_{3pix}$/counts$_{0.5pix}$ $<$ 13 and those in the WF chips
with counts$_{3pix}$/counts$_{0.5pix}$ $<$ 10 were point sources and
thus real GC candidates.  Using these criteria (and confirming the
results with visual inspection), we found that two of the 16 objects
were actually galaxies.

To calculate the surface density of background galaxies in the GC
sample, we first scaled the HST images to the same pixel scale as the
WIYN images, aligned them to the WIYN pointing, and computed the total
area covered by HST.  The HST frames covered 4.41 square arc~minutes
around NGC~7814, yielding a density of
0.45 galaxies per square arc minute.

\subsubsection{Stars}
\label{section:stellar contamination}

We used the latest version of the Galactic structure code from Mendez
and van Altena (1996) and Mendez et al.\ (2000) to estimate the amount
of stellar contamination in the GC sample.  The model allows the user
to choose such parameters as the contribution to star counts from the
Galaxy disk, thick disk, and halo, and the galactocentric distance and
z-height of the Sun.  Output includes the surface density of stars
expected within a given magnitude and color range in a given direction
on the sky.  A selection in two colors was not easily implemented so
we used only a $B-V$ cut.  The model predicts that in the direction of
NGC~7814, the number density of stars with $V$ magnitude and $B-V$
color in the range of our GC candidates is 0.11 stars per square arc
minute.  We varied the Galaxy parameters and the Sun's location and
found that the answer was the same each time.  Applying both a $B-V$
and $V-R$ cut to our source list reduces the stellar contamination
even more than selecting in just one color, so this is likely to be an
overestimate of the stellar contamination.

\subsubsection{Radially-Dependent Contamination Correction}

The fraction of contaminating objects in the GC sample increases with
radial distance as the GC surface density falls off.  To determine
exactly how the contamination fraction varies radially, the 42 GC
candidates in the WIYN sample were assigned to annuli of width
1$\arcm$, starting at the galaxy center and continuing to 7$\arcm$.
The number of non-GCs expected in each annulus was calculated by
multiplying the number density of contaminating objects (0.56 per
square arc minute) by the effective area of the annulus (the portion
where GCs could be detected; see Section~\ref{section:radial
profile}).  Finally, the contamination fraction in each annulus was
calculated by dividing the number of contaminating objects by the
number of GC candidates.

\subsection{Determining the GCLF Coverage}

An observed luminosity function (LF) was created by assigning the $V$
magnitudes of the 42 WIYN GC candidates to 0.3-mag-wide bins.  The
radially-dependent contamination correction was applied to the LF
data.  The total completeness in each LF bin was calculated using the
method described in Paper~I (i.e., the completenesses in all three
filters are convolved to come up with a total value), and the number
of GCs in each bin was divided by the total completeness to produce a
corrected LF.

The sample of 42 GC candidates is too small for use in determining
NGC~7814's intrinsic GCLF.  Instead we assumed the GCLF was a Gaussian
with a given peak magnitude and dispersion and fitted this to the
data, varying only the normalization of the function.  Based on the
GCLF of the Milky Way, we used a peak magnitude of $M_V$ $=$ $-$7.33
(e.g., Ashman \& Zepf 1998), which translates to $V$ $=$ 23.27 at the
distance of NGC~7814.  We fitted Gaussians with dispersions of 1.2,
1.3, and 1.4 magnitudes to the observed data.  The fractional coverage
fitting to the three Gaussians ranged from 0.41 to 0.46, with a mean
value of 0.43$\pm$0.02.

Because there were so few GC candidates and we were fitting the LF
data in the form of a histogram, we investigated the effect of the bin
size on the fitting results.  Changing the bin size can change the
calculated fractional coverage by up to 4 percent.  In
Section~\ref{section:spec freq} we quantify the effect of this and
other uncertainties on the final numbers for NGC~7814's GC system.

\subsection{HST Sources in the Galaxy Disk and Inner Halo}
\label{section:hst sources}

The HST halo pointing was observed in three filters and was therefore
useful for detecting GCs independently from the ground-based data.
This provides a check on the WIYN results and gives radial coverage
into the galaxy disk, where better resolution is especially useful for
finding GCs.  The halo pointing covers part of the disk and extends
$\sim$2$-$3$\arcm$ into the halo.

GC candidates were located in the F555W and F814W images of the halo
pointing using the method described in \citet{kundu99}.  DAOFIND was
used to detect sources above a modest threshold (3 counts above the
background).  Aperture photometry was then performed on each detected
source and a signal-to-noise ratio (SNR) was computed using the counts
in the aperture and the RMS scatter of the background.  A total of 169
point-like objects were selected that had SNR $>$ 2.5 in both images.
Aperture photometry at the locations of the 169 sources was performed
in all three filters using radii of 3 pixels for the PC chip and 2 for
the WF chips and background annuli from 5 to 8 pixels. Zero-points and
aperture corrections for GCs at the distance of NGC~7814 were adopted
from \citet{kw01} to calculate $B$, $V$, and $I$ magnitudes for each
object.

\begin{figure}
\plotone{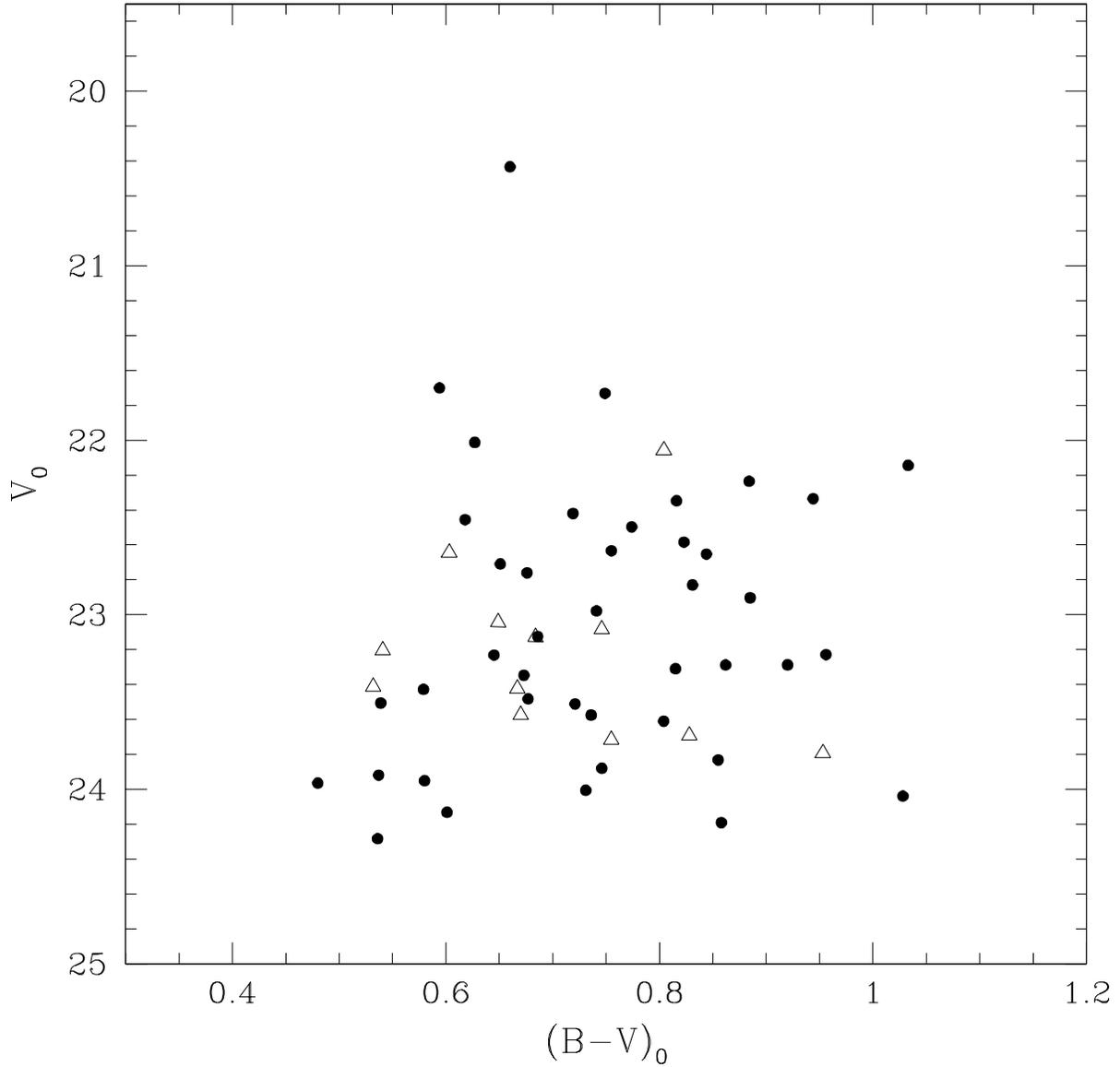}
\caption{Color-magnitude diagram for GC candidates in NGC~7814.  $V$
versus $B-V$ is plotted for 54 GC candidates.  Candidates detected in
the HST images are shown as open triangles.  Candidates from the WIYN
images, or from both the HST and WIYN data, are shown as filled
circles.  Note that the magnitudes and colors have not been corrected
for internal absorption in NGC~7814.\protect\label{fig:color mag}}
\end{figure}

To construct a sample of GC candidates, sources with $V$ magnitudes
between 19.6 and 23.8 were selected.  The faint magnitude cut was
applied so the HST source list would be of similar depth to the WIYN
list.  It also eliminates the need for contamination corrections to
the HST sample, since for magnitudes brighter than $V$ $\sim$ 24 the
contamination level from background galaxies that would be unresolved
in the HST images is negligible \citep{kundu99}.  As before, we chose
an acceptable $B-V$ range of 0.56 to 0.99 for GCs.  Then we selected
objects having $B-V$ and $V-I$ colors and photometric errors that put
them within 3-sigma of the $BVI$ color-color relation for Milky Way
GCs.  The slightly relaxed color criterion (3-sigma vs.\ 2.5 for the
WIYN color selection) allowed for any differences between the HST and
$BVI$ filters.  We also visually inspected each source; several that
turned out to be diffuse knots of emission in the galaxy disk were
eliminated.  The final list of HST GC candidates includes 23 objects.
Eleven of the objects also appeared in the WIYN list and the rest were
not detected by WIYN, e.g., because they were in a masked region near
the galaxy disk or were below the WIYN detection threshold.  A
color-magnitude diagram showing the $V$ magnitudes and $B-V$ colors of
all the GC candidates is shown in Figure~\ref{fig:color mag}.

Simulations following the method described in \citet{kw01} were run to
establish the completeness level of the HST images.  Briefly, a total
of 70,000 artificial point sources with GC-like colors and a range of
magnitudes were added to the $V$ and $I$ images and the detection
steps described above were executed.  The 50\% completeness limits are
$V$ $=$ 25.1 and $I$ $=$ 24.1.

The GCLF fitting and fractional coverage calculation were executed in
the same way as for the WIYN sample.  The resultant GCLF coverage
ranged from 0.48 to 0.53, with a mean value of 0.51$\pm$0.03.

\section{Results}
\label{section:results}
\subsection{Radial Distribution}
\label{section:radial profile}

To construct a radial distribution of GCs, the WIYN GC candidates were
binned into annuli 1$\arcm$ wide, from the center of the galaxy out to
7$\arcm$.  Portions of the WIYN images where GCs could not be reliably
detected --- namely, near the galaxy disk where the noise level is
high and around bright stars where the CCD pixels are saturated ---
were masked out.  An effective area was calculated for each annulus,
which is equal to the area of the annulus minus the masked regions and
the portion extending beyond the edges of the image.

The HST images used to detect GCs covered a smaller area around the
galaxy than the WIYN frames, so to make full use of the data we binned
the GCs into narrower annular regions.  The 23 HST GC candidates were
assigned to 0.5$\arcm$-wide annuli from 0 to 2$\arcm$.  A small
portion of the images, coinciding with a dust lane in the galaxy disk,
was masked out.  Effective areas were calculated in the same way as
for the WIYN data.

To correct the WIYN data for contamination, the number density of
contaminating stars and galaxies was subtracted from the number of GC
candidates in each annulus. The contamination level in the HST sample
was assumed to be insignificant, so no correction was applied.  The
number of GCs in each bin was then divided by the fractional GCLF
coverage (0.43 for the WIYN data and 0.51 for the HST points).
Finally, the radial profile was constructed by calculating a surface
density of GCs for each annulus.  The error on the surface density
takes into account Poisson errors on the number of GCs and
contaminating objects.

\begin{figure}
\plotone{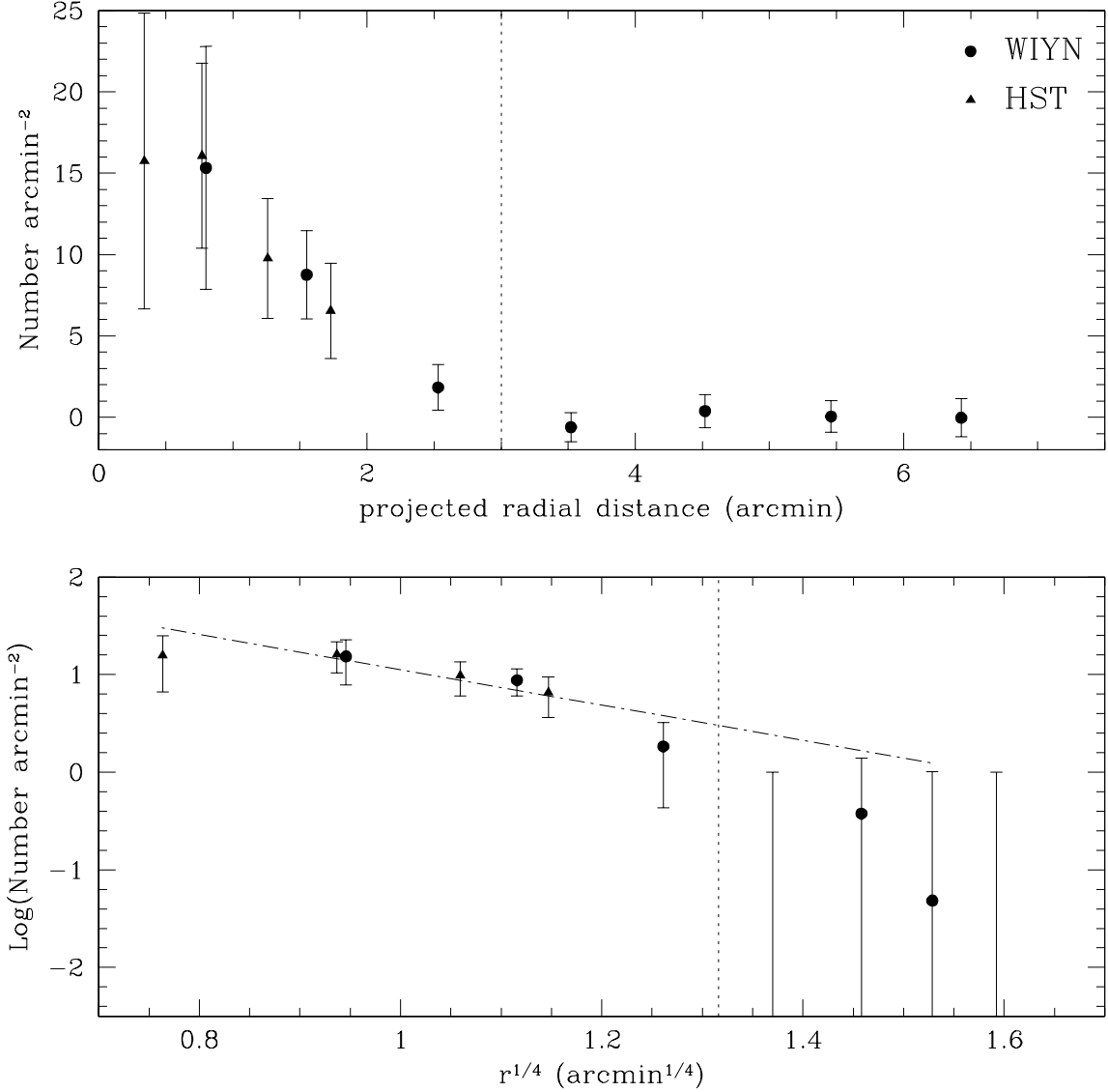}
\caption{Radial distribution of GCs in NGC~7814.  The top figure shows
the surface density of GCs as a function of projected radial distance
from the galaxy center.  The bottom plot shows the log of the surface
density versus $r^{1/4}$ and the dashed line is the best-fit
deVaucouleurs profile.  The vertical dotted line marks 3$\arcm$ in
both plots, beyond which the GC surface density is consistent with
zero within the errors.  The data have been corrected for
contamination, areal coverage, and magnitude incompleteness, as
described in Section~\ref{section:radial profile}.
\protect\label{fig:radial profile}}
\end{figure}

Because in some cases significant portions of the annuli were masked
out or not observed (e.g., close to the galaxy disk and places where
the annuli extend off the images), we calculated the mean distance of
the unmasked pixels in each annulus and used that value to construct
the profile.  The profile is shown in Figure~\ref{fig:radial profile}
and tabulated in Table~\ref{table:radial profile}.  In the figure,
filled circles are the WIYN points and open triangles are from HST.
The top plot shows GC surface density versus radial distance in
arc~minutes and the bottom plot is the log of the surface density
versus radius to the 1/4 power.  The best-fit deVaucouleurs law is
shown as a dashed line in the bottom plot and has the form
log~$\sigma_{GC}$ $=$ (2.86$\pm$0.46) $-$ (1.81$\pm$0.44)~$r^{1/4}$,
where $\sigma_{GC}$ is the surface density of GCs in number per sqare
arc minute and $r$ is the projected radius.  We also fitted a power
law to the data; the best-fit line has the form log~$\sigma_{GC}$ $=$
(1.03$\pm$0.06) $-$ (0.98$\pm$0.25)~log~$r$.  Both the deVaucouleurs
law and power law fit the data fairly well inside 3$\arcm$,
intersecting (or nearly intersecting) the data points plus error bars.
Beyond 3$\arcm$, the deVaucouleurs law is a better match to the data;
in this region, the points and error bars fall systematically below
both of the fitted lines, but the discrepancy is not as large for the
deVaucouleurs law.

For the 3$-$4$\arcm$ annulus and beyond, the GC surface
density is consistent with zero within the errors.  This result
illustrates the importance of obtaining wide-field data when studying
GC systems.  The low values for the GC surface density at radii of
3$\arcm$ and beyond are critically important for constraining the
total number of GCs around NGC~7814, since the inner points can be fit
by many profiles that have quite different outer profiles and
therefore total number.  The observation of very low or zero GC
surface density in the outer annuli strongly suggests that we have
observed the entire radial extent of this galaxy's GC system.  For the
SBF distance modulus, 3$\arcm$ equals 11.5 kpc.  For comparison, we
note that if we projected the Milky Way's GC system onto the $Y$-$Z$
plane (where $Y$ is the Galactocentric coordinate in the direction of
Galactic rotation and $Z$ is the height above or below the plane) and
calculated a radial distance for each GC, 82\% of them would have
radial distances of 11.5~kpc or less (Harris 1996).

\subsection{Total Number and Specific Frequency}
\label{section:spec freq}

The information contained in the corrected GC radial profile can be
used to calculate the total number of GCs around NGC~7814.  Since we
have combined HST data with wider-field WIYN data, we can directly
determine the total number of GCs without the extrapolation from
smaller radius that would have been necessary had we used HST data
alone.  There are two ways to compute the total number: one is to
integrate the best-fit deVaucouleurs profile from $r$ $=$ 0 to an
outer radius, and the other is to sum the actual data points, i.e., by
multiplying the surface density at each point in the profile by the
area of the associated annulus and summing over all radii.  Both
methods produce a total number of GCs for the galaxy, corrected for
magnitude incompleteness, missing spatial coverage, and contamination
from non-GCs.  The best-fit deVaucouleurs law provides a good fit to
the radial profile data between $\sim$0.8$\arcm$ and 1.7$\arcm$, and
either slightly or significantly overestimates the data elsewhere.
For this reason we have used both methods --- integrating the profile
and summing the actual data --- to calculate the total number of GCs
in NGC~7814.  Because the GC surface density is consistent with zero
from 3$\arcm$ outward, and the deVaucouleurs function consistently
overestimates the data points beyond that radius, we stop the
integration or summation at that point.  (Below we discuss the
possible impact of our choice of integration limit on the final
results.)  Summing the data points in the profile to 3$\arcm$ yields a
total of 140 GCs for NGC~7814.  Integrating the best-fit deVaucouleurs
law from 0 to 3$\arcm$ yields 190 GCs.

The total number of GCs can be normalized by luminosity or mass of the
galaxy in order to facilitate comparison of NGC~7814's GC system to
that of other galaxies.  The specific frequency, $S_N$, is the number
of GCs normalized by luminosity and is defined as
\begin{equation}
{S_N \equiv {N_{GC}}10^{+0.4({M_V}+15)}}
\end{equation}
\citep{hvdb81}.  The total, extinction-corrected, face-on magnitude
for NGC~7814 is $V^T_0$ $=$ 10.20~(RC3; deVaucouleurs et al.\ 1991).
Since $R_{25}$ for this galaxy is 2.75$\arcm$, virtually all of the
galaxy light is located within $\sim$3$\arcm$.  Therefore, integrating
the GC profile to 3$\arcm$ and then using the total luminosity to
calculate $S_N$ is a reasonable approach. Assuming the SBF distance
modulus yields a total absolute magnitude of $M^T_V$ $=$ $-$20.40.
Combining this with the total numbers of GCs from summing the data or
integrating the deVaucouleurs profile yields, respectively, $S_N$ $=$
1.0 or 1.3.  

The primary source of uncertainty on $S_N$ is likely to be the galaxy
magnitude, particularly since NGC~7814 is an edge-on spiral and
correcting for extinction in such cases is difficult.  The error on
the total $V$ magnitude quoted in RC3 is 0.13~mag, so $M^T_V$ is at
least that uncertain.  If we assume that the actual total magnitude of
NGC~7814 may be up to 0.3~mag brighter than the RC3 value, the result
is a decrease in $S_N$ of $\sim$0.3.  Another source of uncertainty is
associated with the GCLF coverage calculation.  Depending on the
dispersion assumed for the GCLF, the estimated coverage changed by as
much as 5\%.  The coverage also varied by 4\% depending on the bin
size of the data.  Taking both of these into account, the fractional
coverage is known to $\sim$6\%.  This translates to an uncertainty in
$S_N$ of $\sim$0.1.  The Poisson errors on the total numbers of GCs
and contaminating objects results in an additional 0.1 uncertainty.
Finally, there is uncertainty associated with the integration of the
radial profile.  One effect is that we stopped the integration at
3$\arcm$ (11.5~kpc), beyond which the GC surface density drops to
zero.  As mentioned, $\sim$18\% the Milky Way's GCs are located at
radial distances beyond 11.5~kpc and it is possible that we have
missed some portion of NGC~7814's GC system at these larger radii.
Increasing the total number of GCs in NGC~7814 by 18\% would increase
$S_N$ by $\sim$0.2.  A second effect is that using a direct summation
to 3$\arcm$ produces a $S_N$ value 0.3 lower than integrating the
de~Vaucouleurs fit over the same region.  If we combine the above
sources of uncertainty on $S_N$ in quadrature, we obtain a final value
of $S_N$ of 1.3$\pm$0.4 for NGC~7814.

Normalizing the number of GCs by host galaxy mass makes it easier to
compare the GC systems of galaxies with different star formation
histories (such as spirals and ellipticals) and thus different
mass-to-light ratios in their stellar populations.  Zepf \& Ashman
(1993) suggest using a parameter called $T$, which they define as
\begin{equation}
T \equiv \frac{N_{GC}}{M_G/10^9\ {\rm M_{\sun}}}
\end{equation}
where $N_{GC}$ is the number of GCs and $M_G$ is the mass of the host
galaxy.  Following \citet{za93}, we adopt a mass-to-light ratio
$M/L_V$ $=$ 6.1 for an Sab spiral like NGC~7814.  Combining this with
$N_{GC}$ $=$ 140 yields $T$ $=$ 1.9, and using $N_{GC}$ $=$ 190 gives
$T$ $=$ 2.5 for NGC~7814.  The values derived for total number and
specific frequency are summarized at the end of this section.

\subsection{Mass-Normalized Number of Blue Globular Clusters}
\label{section:Tblue}

One of the main objectives of our overall study is to test a specific
prediction of the AZ92 merger scenario for the origin of ellipticals.
In this scenario, two (or more) disk galaxies with their own GC
systems merge to form an elliptical.  A metal-rich population of GCs
forms from the colliding gas as the spirals merge.  The metal-poor GCs
in the progenitor spirals become the metal-poor population in the
elliptical.  The end result is an elliptical with bimodal color and
metallicity distributions of GCs.  As mentioned in the Introduction, a
consequence of this scenario is that the mass-normalized number of
metal-poor GCs in a typical spiral should be the same as in
ellipticals.  In this simple merger picture, the GCs created in the
merger are formed from enriched gas (and are thus metal-rich), so even
if a given elliptical has been formed from multiple disk galaxies, the
mass-normalized number of metal-poor GCs ($T$ for the metal-poor
population, which we will refer to as $T_{\rm blue}$) should stay
fairly constant.  We can directly test the AZ92 prediction by
comparing $T_{\rm blue}$ for NGC~7814 (assuming it to be a typical Sab
spiral) to typical $T_{\rm blue}$ values for ellipticals.

Table~\ref{table:spec freq} shows that $T$ for NGC~7814's {\it entire}
GC system is within the range 1.9 to 2.5.  To determine $T_{\rm blue}$
we first need to select a subsample of GC candidates that is equally
complete in all three filters.  The reddest GC candidate in our list
has $B-R$ $=$ 1.64.  The GC list is 90\% complete at $B$ $=$ 24.3, so
to construct a ``complete sample'', we cut the 42-candidate list at
$R$ $=$ 22.7.  This yields 21 objects.  One of the objects is extended
in the HST images, so this leaves 20 objects.

Given sufficient numbers of GC candidates, one can derive robust
values for the relative proportions of red and blue GCs in a galaxy.
This is typically done by fitting multiple Gaussian functions to a
color distribution made with a complete sample, e.g., using the KMM
algorithm \citep{abz94}.  We did this in Paper~I using 366 GC
candidates in NGC~4472 and found that the color distribution was
significantly bimodal, with a gap between the blue and red populations
at $B-R$ $=$ 1.23.  The sample of 20 objects in NGC~7814 does not
appear bimodal
and is too small to use with KMM, which requires at least 50 objects
to produce a reliable result.
If we instead simply divide the sample in the way that NGC~4472's GC
population is divided (i.e., at $B-R$ $=$ 1.23), we find that eight GC
candidates are blue and 12 are red. With such small numbers the
uncertainties on the overall proportions of blue and red GCs in the
system are large.
Moreover, there are possible systematic errors in play: some of the GC
candidates are likely reddened due to internal extinction in NGC~7814,
in which case more may actually belong in the blue category.  (In
fact, six of the 13 GCs with $B-R$ $>$ 1.23 have radial distances
within 2$\arcm$ of the center of NGC~7814, where the diffuse galaxy
light is appreciable; reddening could very well be a factor for some
of these objects.)  Because of these uncertainties, we will adopt 40\%
as a {\it lower limit} on the percentage of blue GCs in the NGC~7814
system.  Applying this percentage to $T$ $=$ 2.5 yields a
mass-normalized number of blue GCs, $T_{\rm blue}$, of 1.0.

Even if several of the GC candidates in the complete sample are
affected by reddening, it is unlikely that {\it all} of them actually
belong in the blue category.  To estimate an upper limit for the
proportion of blue GCs, we look at the most well-known spirals, the
Milky Way and M31.  About 70\% of the GCs in the Milky Way are part of
the metal-poor, ``halo'' population (Harris 1996; C\^ot\'e 1999).  The
fraction is similar in M31, at 66\% \citep{barmby00}. If we assume
that the proportion of blue GCs in NGC~7814 is also $\sim$70\%, and
that the total $T$ is 2.5, then $T_{\rm blue}$ for this galaxy is 1.8.

By the above arguments, $T_{\rm blue}$ for NGC~7814 is between $\sim$1
and 2.  We will compare this range of values to that of other galaxies
and discuss the implications for galaxy formation in
Section~\ref{section:discussion}.

\subsection{Effect of Change in Distance on Specific Frequency}
\label{section:distance effect}

Since distances to individual galaxies have some uncertainty, we have
explored the effect of changing the galaxy distance on the total
number and specific frequency of GCs.  The SBF distance used in this
paper (m$-$M $=$ 30.60$\pm$0.14, or 13.2 ~Mpc) was only recently
published \citep{tonry01}.  Despite being an edge-on spiral, NGC~7814
does not have a published Tully-Fisher distance, so B92 used its
radial velocity and a Virgocentric flow model to derive a distance of
12.5~Mpc (m$-$M $=$ 30.48).  They also calculated results using a
distance modulus 1 mag further, based on the fact that their GCLF
changed appreciably depending on whether they treated one of their
observed fields as background or GC system.

To illustrate the effect of distance on the derived properties of
NGC~7814's GC system, we will adopt a distance modulus of 30.95, which
is the SBF value plus 2.5-$\sigma$.  The larger distance increases the
total number of GCs, primarily because the GCLF coverage is reduced,
but the effect on $S_N$ is counteracted by a brighter absolute
magnitude for the galaxy.  The luminosity- and mass-normalized GC
specific frequencies for NGC~7814 at this alternative distance are
$S_N$ $=$ 0.8 and 1.0 (from summing the data and integrating the fit)
and $T$ $=$ 1.4 and 1.9 (data and fit).  Therefore, using a farther
distance results in a slightly smaller specific frequency.

Table~\ref{table:spec freq} lists the total numbers and specific
frequencies for NGC~7814's system for the two different distance
moduli.  Columns (1) and (2) give the assumed distance modulus and
distance to the galaxy and column (3) the galaxy luminosity.  Columns
(4)---(6) are, respectively, the total number of GCs, the
luminosity-normalized specific frequency $S_N$ and the mass-normalized
specific frequency $T$ for the total system.  The first number given
in each of columns (4)---(6) comes from summing the data points in the
radial profile and the second (larger) number from integrating the
best-fit deVaucouleurs function.

\section{Discussion}
\label{section:discussion}

As stated in the Introduction, NGC~7814 was of particular interest to
us as a possible example of a spiral galaxy with an extremely populous
GC system.  When we began our study, it was thought to have by far the
largest GC specific frequency of any spiral \citep{harris91,az98}.  We
find instead that NGC~7814 has $_<\atop{^\sim}$200 GCs and
$S_N$~$\sim$~1.  This makes it comparable to our own Galaxy, which has
$\sim$150 known GCs, an estimated total population of $\sim$180 and
$S_N$ $\sim$ 0.5$-$0.7, and M31, with $N_{GC}$ estimated at $\sim$450
and $S_N$ $\sim$ 0.9$-$1.3 \citep{az98, barmby03}.

In their study of NGC~7814, B92 detected $\sim$120 GC candidates and,
correcting for areal and magnitude incompleteness and extrapolating
the radial profile to both smaller and larger radius, concluded that
it had a GC system with 500$-$1000 GCs.  They used the technology
available at the time, namely a CCD with 0.4$\arcs$-pixels, which made
it more difficult to eliminate contaminating galaxies.  Furthermore,
they observed the galaxy in a single filter ($V$), so selecting GCs by
color was not possible.  They observed several fields around the
galaxy to cover a substantial fraction of the GC system, as well as
control fields away from the target to help correct for contamination.
The latter should in theory work well, but as noted in
Section~\ref{section:extended source cut} there appears (from our
data) to be an overdensity of background galaxies in the direction of
NGC~7814.  If their control field did not have a correspondingly high
density of background objects, they would have under-corrected for
contamination, resulting in an overestimate of the number of GCs in
NGC~7814.  We note that for the Virgo elliptical NGC~4472, our
techniques for reducing and correcting for contamination also resulted
in a smaller $S_N$ value (Paper~I) than previous estimates.

A major objective of our GC system survey is to test whether the
mass-normalized total numbers of metal-poor GCs in spirals and
ellipticals are the same, as expected, for example, in the AZ92 merger
model for the formation of ellipticals.  The numbers of GCs in
ellipticals vary with the luminosity of the host galaxy, with more
luminous galaxies generally hosting larger populations of globulars
(e.g., Harris \& Racine 1979, Djorgovski \& Santiago 1992, Zepf,
Geisler, \& Ashman 1994).  Therefore to fully test the AZ92 picture we
need to compare $T_{\rm blue}$ for spirals with ellipticals of varying
luminosities.  As part of our survey, we are studying the total GC
populations of a sample of early-type galaxies with a range of
luminosities.  An example of a modest-luminosity elliptical is
NGC~3379, an E1 with $M_V$ $\sim$ $-$21.  We find that this galaxy has
$\sim$270 GCs.  Using the above $V$-band magnitude and a mass-to-light
ratio $M/L_V$ of 10 for ellipticals \citep{za93}, this translates to a
total $S_N$ of 1.2 and $T$ $\sim$ 1.4 (Rhode \& Zepf, in preparation;
Paper III).  Approximately 70\% of the GC candidates in NGC~3379 are
blue, so $T_{\rm blue}$ $\sim$1.  By comparison, in
Section~\ref{section:Tblue} we estimated that $T_{\rm blue}$ for
NGC~7814 is between 1 and 2.  In the Milky Way and Andromeda, about
70\% of the GCs are metal-poor, so for these galaxies $T_{\rm blue}$
is, respectively, $\sim$0.9 and $\sim$1.2.  These numbers indicate
that it is possible to create the blue GC population of NGC~3379
simply by merging the GC systems of spirals like NGC~7814 or the Milky
Way.  Thus it appears that --- in terms of the total numbers of GCs
--- a simple merger scenario can account for the GC system of a galaxy
like NGC~3379 if one assumes, as AZ92 did, that the number of new GCs
formed in the merger is roughly equal to the number originally
contained in the spirals.

NGC~4472 is the brightest galaxy in Virgo (with $M_V$ $\sim$ $-$23)
and is a good example of a giant cluster elliptical.  In Paper~I, we
calculated a total number of GCs ($\sim$5900) and the proportions of
red and blue GCs in NGC~4472 (40\% and 60\%, respectively).  Again
assuming a mass-to-light ratio $M/L_V$ of 10, we derived $T$ $=$ 2.6
for the blue, metal-poor GC population of NGC~4472.  While $T_{\rm
blue}$ for NGC~7814 is larger than for the Milky Way and M31, at
between $\sim$1 and 2, it does not appear to be large enough to fully
account for the large blue GC population in a galaxy like NGC~4472.
In this case our data suggest that a simple merger scenario along the
lines of AZ92 cannot explain how massive ellipticals like NGC~4472 and
their GC systems formed.

To reiterate, the results from our study so far indicate that the
metal-poor GC populations of modest-luminosity, low-$S_N$ ellipticals
can be accounted for by the merger of typical spirals, whereas $T_{\rm
blue}$ for a luminous, high-$S_N$ elliptical like NGC~4472 is larger
than for any spiral studied to date.  We note that other authors
(e.g., Harris 1981; van den Bergh 1984; Harris 2001) have argued
similarly, that the GC systems of giant cluster Es cannot have formed
via simple spiral-spiral mergers.  These previous arguments were
typically based on comparisons of total $S_N$ for spirals and
ellipticals rather than the specific frequency of the blue GCs alone.
A final effect that needs to be accounted for in our discussion of $T$
is that mass-to-light ratios of ellipticals vary with galaxy
luminosity.  More luminous ellipticals are redder than
lower-luminosity ones, and for all stellar populations models (e.g.,
Bruzual \& Charlot 1993) this leads to higher stellar $M/L_V$ for
higher-$L$ ellipticals.  We will discuss the implications of this in
more detail in Paper III.  Models of this effect (e.g., Zepf \& Silk
1996; see also Dressler et al.\ 1987) suggest that stellar population
differences between ellipticals of different luminosities can account
for up to about half of the increase in $T$ with increasing elliptical
galaxy luminosity.  Thus, even with variations in $M/L_V$ for
ellipticals, there remain differences between $T_{\rm blue}$ for
spirals (and low-luminosity Es) compared to luminous Es. This seems to
suggest that GC systems form or evolve differently depending on the
luminosity of the host galaxy.

One possibility is that the number of old, metal-poor GCs is higher in
more luminous ellipticals because there is less dynamical destruction
of GCs in these galaxies compared to lower-luminosity ellipticals and
spirals.  \citet{mw97a} modeled the evolution of the GC system of M87
and used scaling relations to translate their results to ellipticals
of various luminosities.  They suggest that ellipticals may have
formed with similar GC specific frequencies but that GCs in
lower-luminosity Es underwent more dynamical destruction due to their
higher densities, resulting in a trend of higher $S_N$ with higher
galaxy luminosity.  Similarly, \citet{vesper00} modeled GC destruction
in ellipticals with a range of masses and found that the fraction of
surviving clusters increases with increasing galaxy mass.  With regard
to spirals, simulations exploring the effect of dynamical destruction
on GCs in the Milky Way system (e.g., Murali \& Weinberg 1997b;
Vesperini 1998) suggest that, just as in low-luminosity ellipticals,
it is likely that a significant fraction of the population has been
destroyed over a Hubble time. Therefore, while this is certainly not
required to be the case, there remains the possibility that spirals at
high redshift had large enough metal-poor GC populations to account
for what we see in luminous ellipticals today.

Recent work on hierarchical merging scenarios for galaxy formation
(e.g., Beasley et al.\ 2002, Santos 2003) provides another possible
explanation for the apparent luminosity-dependence of $T_{\rm blue}$.
In these types of scenarios, metal-poor GCs are formed at high
redshift in protogalactic fragments; the gas-rich fragments merge over
time, producing metal-rich GCs and eventually resulting in the massive
galaxies and GC systems that we see today.  Santos (2003) proposes
that metal-poor GC formation occured at redshifts prior to
$z$~$\sim$~7 before being suppressed by reionization.  In this
picture, the timing of the collapse of structures in the
pre-reionization Universe is what determines how many blue GCs are
contained in a given end-product galaxy.  Massive galaxies like giant
cluster Es collapsed earlier, which means that a larger fraction of
their eventual mass was in place by the time reionization occurred.
Less massive galaxies like spirals and field galaxies collapsed later,
so fewer blue GCs were produced in them before GC formation was
halted.  The natural result is that more massive (and thus more
luminous) galaxies will have more blue, metal-poor GCs per unit mass
than less massive objects.  Within the framework of this scenario, the
observed spatial distributions of the metal-poor GCs can help place
constraints on the redshifts at which the protogalactic fragments
collapsed, as well as on the masses of the fragments.  In hierarchical
models, the most overdense fragments that collapse at the highest
redshifts (and can begin to form GCs) are more centrally-concentrated
than those with lower densities that collapse later.  Likewise,
fragments with more mass will also collapse earlier and be more
concentrated toward the galaxy center. Santos has simulated this in
detail for the Milky Way GC system, but has not yet produced testable
predictions for the GC spatial distributions of other galaxies.  In
future, however, our observed GC spatial profiles for galaxies with a
variety of masses and in different environments will provide valuable
constraints for this theoretical work.

\section{Summary}
\label{section:summary}

We have undertaken a wide-field CCD survey of the GC systems of a
large sample of spiral and elliptical galaxies with the primary goal
of testing models for the origin of ellipticals, including the idea
that they are formed via spiral-spiral mergers.  We have completed our
study of NGC~7814, an Sab spiral that was found in past work to have
an extremely high GC specific frequency.  Our main results are as
follows:

1. Our data provide radial coverage of NGC~7814's GC system to
   $\sim$7$\arcm$ ($\sim$30~kpc) from the galaxy center.  The surface
   density in the radial distribution of GCs is consistent with zero
   by $\sim$3$-$4$\arcm$ ($\sim$11$-$15~kpc) from the galaxy center,
   suggesting that we have observed the full extent of NGC~7814's
   GC system.

2. NGC~7814 has a total population of $\sim$200 GCs and a
   luminosity-normalized specific frequency $S_N$ of 1.3~$\pm$0.4.
   NGC~7814 is not a high-$S_N$ spiral as previously claimed, but
   instead has a GC system that is comparable to those of the two most
   well-studied spirals, the Milky Way and M31.

3. A natural consequence of the fact that NGC~7814 has many fewer GCs
   than previously thought is that the number of metal-poor GCs must
   also be substantially reduced.  We compare NGC~7814's blue GC
   population to that of ellipticals from our study in order to test a
   prediction of the Ashman \& Zepf merger model.  We find that
   merging the blue GC populations of spirals like NGC~7814 is enough
   to account for the blue GCs in a moderate-luminosity field elliptical
   like NGC~3379, but not in a giant cluster elliptical like NGC~4472.
   This suggests that simple mergers may be able to explain
   the origins of typical-luminosity ellipticals but that some
   additional process is needed to account for the GC systems of more
   massive ellipticals.

Our findings for NGC~7814 significantly alter the overall picture of
extragalactic GC systems. In Paper III of this series, we will present
results for the remaining early-type galaxies in the sample and will
specifically address predictions of the various scenarios for the
formation of ellipticals, including (in addition to the merger model)
multi-phase collapse \citep{forbes97}, collapse plus accretion
\citep{cote98}, and hierarchical merging (e.g., Beasley et al.\ 2002,
Santos 2003).



\acknowledgments

K.L.R. gratefully acknowledges financial support from a NASA Graduate
Student Researchers Fellowship for this project. S.E.Z. acknowledges
support for this work from the Michigan State University Foundation
and NASA Long-Term Space Astrophysics Grant NAG5-11319.  We are
grateful to Arunav Kundu for his assistance with the HST data
analysis, which included providing a list of detected objects in the
HST halo pointing and running completeness tests on those data.  We
thank the staff at WIYN and Kitt Peak National Observatory for
assistance during our observing runs.  We also thank the anonymous
referee for providing useful comments that improved the quality of the
paper.  This research has made use of the NASA/IPAC Extragalactic
Database (NED), which is operated by the Jet Propulsion Laboratory,
California Institute of Technology, under contract with the National
Aeronautics and Space Administration.


\clearpage

%
%

\clearpage

\clearpage

\begin{deluxetable}{lr}
\tablecaption{Aperture Corrections Used for Photometry of WIYN Sources}
\tablewidth{170pt}
\tablehead{\colhead{Image} & \colhead{Aperture Correction}}
\startdata
$B$ & $-$0.223 $\pm$ 0.006\\
$V$ & $-$0.179 $\pm$ 0.006\\
$R$ & $-$0.230 $\pm$ 0.005\\
\enddata
\tablecomments{The values listed here are the difference
between the total magnitude and the magnitude within an aperture of
radius equal to the average FWHM of the image.}
\protect\label{table:apcorr}
\end{deluxetable}

\clearpage
\begin{deluxetable}{lr}
\tablecaption{Corrected Radial Profile for GCs in NGC~7814}
\tablewidth{160pt}
\tablehead{\colhead{Radius} & \colhead{Surface Density}\\
\colhead{(arcmin)} & \colhead{(arcmin$^{-2}$)}}
\startdata
0.34 &  15.76 $\pm$   9.10  \\
0.77 &  16.07 $\pm$   5.68  \\
0.80 &  15.33 $\pm$   7.47  \\
1.26 &   9.77 $\pm$   3.69  \\
1.55 &   8.76 $\pm$   2.70  \\
1.73 &   6.55 $\pm$   2.93  \\
2.53 &   1.84 $\pm$   1.40  \\
3.52 &  $-$0.61 $\pm$   0.90\tablenotemark{\dag} \\
4.52 &   0.38 $\pm$   1.02  \\
5.46 &   0.05 $\pm$   0.96  \\
6.43 &  $-$0.03 $\pm$   1.17\tablenotemark{\dag}  \\
\enddata
\tablenotetext{\dag}{A contamination correction has been
applied to the surface density of GCs in each bin, resulting in a
negative surface density in two of the outer bins.}
\protect\label{table:radial profile}
\end{deluxetable}

\clearpage
\begin{deluxetable}{lclllll}
\tablecaption{Total Numbers and Specific Frequencies for NGC~7814's GC
System}
\tablewidth{330pt}
\tablehead{\colhead{$m-M$} & \colhead{Distance} & \colhead{$M_V^T$}
& \colhead{$N_{GC}$} & \colhead{$S_N$}& \colhead{$T$} \\
\colhead{} & \colhead{(Mpc)} & \colhead{} & \colhead{} & \colhead{} & \colhead{}}
\startdata
30.60 &	13.2 &	$-$20.40 & 140$-$190 & 1.0$-$1.3 & 1.9$-$2.5\\
30.95 &	15.5 &  $-$20.75 & 150$-$200 & 0.8$-$1.0 & 1.4$-$1.9\\
\enddata
\protect\label{table:spec freq}
\end{deluxetable}

\end{document}